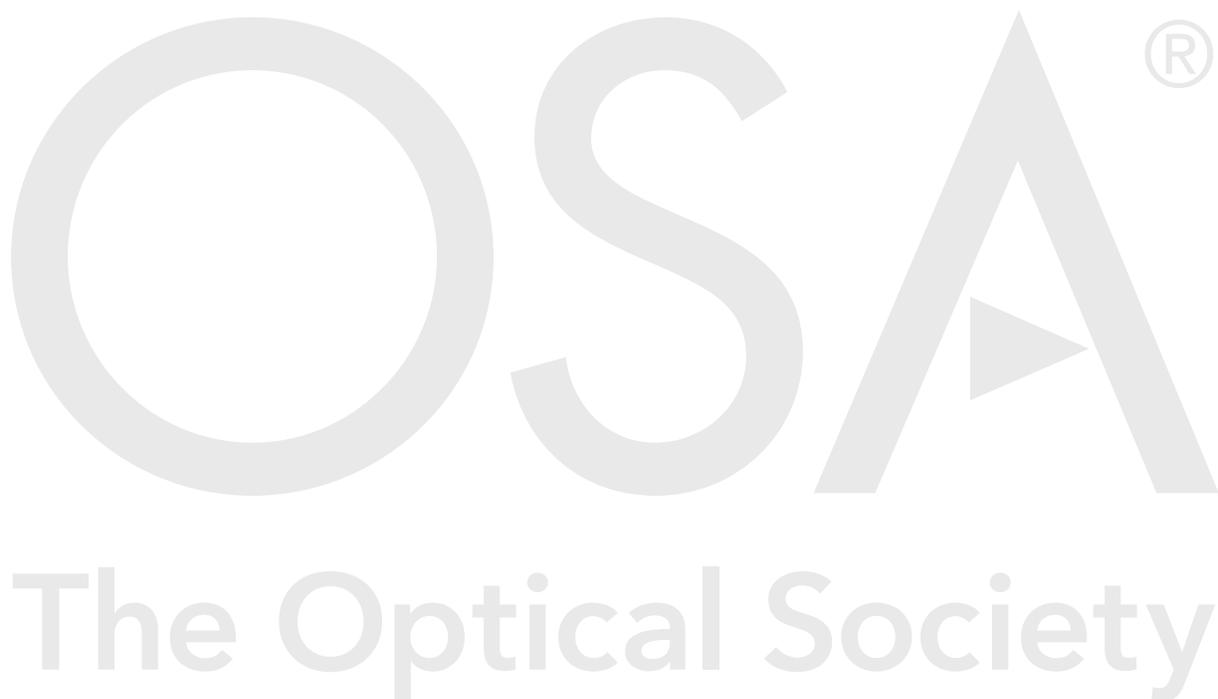

# Wavelength-switchable ultra-narrow linewidth fiber laser enabled by a figure-8 compound-ring-cavity filter and a polarization-managed four-channel filter

**TING FENG,[1] DA WEI,[1] WENWEN BI,[1] WEIWEI SUN,[1] SHENGBAO WU,[1] MEILI JIANG,[1] FENGPING YAN,[2] YUPING SUO,[3] AND X. STEVE YAO[1,4,*]**

[1]*Photonics Information Innovation Center, Hebei Provincial Center for Optical Sensing Innovations, Hebei University, Baoding 071002, China*
[2]*School of Electronic and Information Engineering, Beijing Jiaotong University, Beijing 100044, China*
[3]*Shanxi Provincial People's Hospital, Shanxi Medical University, Taiyuan 030012, China*
[4]*NuVision Photonics, Inc., Las Vegas, NV 89109, USA*
*\*syao@ieee.org*

**Abstract:** We propose and demonstrate a high-performance wavelength-switchable erbium-doped fiber laser (EDFL), enabled by a figure-8 compound-ring-cavity (F8-CRC) filter for single-longitudinal-mode (SLM) selection and a polarization-managed four-channel filter (PM-FCF) for defining four lasing wavelengths. We introduce a novel methodology utilizing signal-flow graph combined with Mason's rule to analyze a CRC filter in general and apply it to obtain the important design parameters for the F8-CRC used in this paper. By combining the functions of the F8-CRC filter and the PM-FCF assisted by the enhanced polarization hole-burning and polarization dependent loss, we achieve the EDFL with fifteen lasing states, including four single-, six dual-, four tri- and one quad-wavelength lasing operations. In particular, all the four single-wavelength operations are in stable SLM oscillation, typically with a linewidth of <600 Hz, a RIN of $\leq$−154.58 dB/Hz@$\geq$3 MHz and an output power fluctuation of $\leq \pm3.45\%$. In addition, all the six dual-wavelength operations have very similar performances, with the performance parameters close to those of the four single-wavelength operations, superior to our previous work and others' similar work significantly. Finally, we achieve the wavelength-spacing tuning of dual-wavelength operations for photonic generation of tunable microwave signals, and successfully obtain a signal at 23.10 GHz as a demonstration.

## 1. Introduction

Single-longitudinal-mode (SLM) erbium-doped fiber laser (EDFL) in C-band (1530–1565 nm) has been adopted or anticipated as the preferred light source for many important applications, due to its outstanding inherent merits such as low intensity and phase noise, narrow linewidth, high beam quality, and excellent compatibility with fiber optical systems. Single-wavelength EDFLs with narrow linewidth and high stability have been utilized commercially or tested experimentally in the cutting-edge applications such as ultra-long distance coherent optical communication, coherent fiber optic sensing, high resolution optical metrology and spectroscopy, and coherent Doppler LiDAR [1-5]. They are also promising in the potential applications relating to optical atomic clocks, measurements of fundamental constants and physics [6, 7]. A multi-wavelength EDFL (MW-EDFL), with narrow linewidth SLM operation for every lasing wavelength, is strongly desired as the light source in photonic generation of high spectral purity microwave signals, multi-parameter optical fiber sensing, and wavelength division multiplexing fiber systems in the future [8-12]. Therefore, it is attractive to achieve flexible switching among single- and multi-wavelength operations in a single SLM EDFL system. Generally, a switchable MW-EDFL with stable SLM oscillation for every lasing line involves the following key techniques: I) SLM selection, II) multi-channel filter to define the primary lasing wavelengths, III) switching mechanism to control the operation state, and IV) mechanism to stabilize the multi-wavelength operation.

Ultra-short distributed feedback and distributed Bragg reflector (DBR) oscillating structures are the easiest and most mature methods to achieve SLM operation for a fiber laser [13-16]. However, the limited power, low flexibility and difficulty for achieving multi-wavelength operation are the major shortcomings. A long oscillating cavity with an ultra-narrow filter to obtain the SLM operation is another main method [17, 18]. However, the severe spatial-hole-burning induced by the standing-wave effect in gain fibers, leading to an unstable SLM lasing, cannot be avoided in the DBR based and other linear cavities. In contrast, a long ring structure with an ultra-narrow filter is a preferable scheme for the SLM EDFL, especially for a switchable MW-EDFL. Commonly available ultra-narrow optical filters include the high finesse Fabry–Pérot (F-P) etalon [19] and the fiber Bragg grating (FBG) based filters, such as phase-shifted FBG [20], FBG-based F-P filter [21], and chirped moiré FBG [22], but they are all complicated or expensive. Compound-ring-cavity (CRC) made of several cascaded or nested fiber rings is a low-cost and easy alternative to obtain high-quality ultra-narrow filters, which has been verified by our group [23-26] and others [27-30]. In the previous work [24], we designed a dual-coupler ring based compound-cavity (DCR-CC) filter with a passband of ~8.20 MHz and a free spectrum range (FSR) of ~10.22 GHz, which has an excellent SLM selection capability and is made with four fiber optical couplers (OCs) with a very low cost of <20 US dollars. In this paper, we will present an even lower cost and more compact CRC filter, named figure-8 CRC (F8-CRC) filter made of only three OCs. This F8-CRC also has the superb capability to enable SLM lasing for the EDFL in this paper. Furthermore, in this work, we introduce a new efficient analysis methodology for the CRC filters dedicated to SLM fiber lasers in general.

There are numerous multi-channel filters proposed for defining primary lasing wavelengths in MW-EDFLs [22, 23, 25, 31-45]. Multiple wavelength-switching mechanisms have been successfully used, such as those based on polarization control [22, 24, 25, 34, 39, 40], nonlinear polarization rotation (NPR) [36, 37], intensity-dependent inhomogeneous loss [46], and tunable-filter scanning [33]. Additionally, to suppress the wavelength competition induced by the homogenous broadening effect of rare-earth-doped gain fibers to stabilize the multi-wavelength lasing, the mechanisms involving the NPR based or nonlinear amplifier loop mirror based amplitude equalizer [25, 47], four-wave mixing [48-50], and polarization-hole-burning (PHB) [23, 24, 44] are reported in literatures. Among the proposed techniques, the polarization control is an excellent way to achieve the switching among different operation states of MW-EDFLs, especially combining it with a polarization-dependent multi-channel filter, such as a high-birefringence FBG (HB-FBG) [23]. The HB-FBG, with two narrow reflection channels, can be fabricated in a segment of polarization maintaining fiber (PMF). Meanwhile, the HB-FBG in a laser cavity can introduce strong PHB effect in the gain fiber [23], which can mitigate wavelength competition obviously to obtain stable dual-wavelength operation. In the previous work [24], we designed a four-channel superimposed HB-FBG (SI-HB-FBG) reflecting filter, and used it in a four-wavelength-switchable EDFL (4WS-EDFL) to obtain switching operation among 15 lasing states, including single-, dual-, tri- and quad-wavelength. However, since the lasers reflected from channel-1 and channel-3, channel-2 and channel-4 of the SI-HB-FBG were respectively with the same state of polarization (SOP), the switching among multi-wavelength operations were a bit difficult and the fluctuations of the two lasers simultaneously oscillating with a same SOP were larger than the two lasers with orthogonal SOPs. In order to resolve these issues, in this work, we propose a novel design to fusion splice two HB-FBGs together with a 45 °polarization-axial-offset to form a novel polarization-managed four-channel filter (PM-FCF), in which the reflected lights from four channels all have different SOPs. In addition, using the SI-HB-FBG as a four-channel filter in the 4WS-EDFL [24], the spacing between any two channels of which cannot be changed. However, using the PM-FCF, the wavelength-spacing between different channels of two HB-FBGs can be changed by easily stretching one of the HB-FBGs while keeping the other one static. To date, there is only one report on a four-channel filter made of two cascaded HB-FBGs for achieving a switchable multi-wavelength fiber laser [51], however the authors used a polarization controller (PC) between two HB-FBGs to rotate the polarization. Not only their approach is complicated and unstable, but also has the issue that the laser outputs are all not in SLM oscillation.

In this paper, in a ring cavity fiber laser we introduce a F8-CRC and a PM-FCF as enabling components for SLM selection and for defining lasing wavelengths, respectively, both for the first time. We achieve a high performance 4WS-EDFL, capable of switching among 15 states, including 4 single-, 6 dual-, 4 tri- and 1 quad-wavelength operations. Specifically, we introduce a novel methodology utilizing the signal-flow graph combined with the Mason's rule to analyze a CRC filter in general, and apply it to obtain the important design parameters for the F8-CRC filter. The expected F8-CRC is fabricated successfully with the designed parameters. Next, we use the enhanced PHB effect introduced mainly by the PM-FCF in a coiled EDF to mitigate wavelength competition. We then control the SOPs to bring polarization dependent loss (PDL) for different lasers, and finally achieve the switchable multi-wavelength operations. As part of the demonstration, in all 4 single- and 6 dual-wavelength operations, the 4WS-EDFL exhibits superb performances in terms of spectrum stability, optical signal to noise ratio (OSNR), SLM, linewidth, output power stability, relative intensity noise (RIN) and polarization characteristics. We also demonstrate the 4WS-EDFL's capability for the photonic generation of tunable microwave signals. Finally, the superiority for tri- and quad-wavelength switchable operations based on the PM-FCF is validated, in comparison with what is achieved with the SI-HB-FBG.

## 2. Experimental setup, principle and theory

### 2.1 Laser configuration

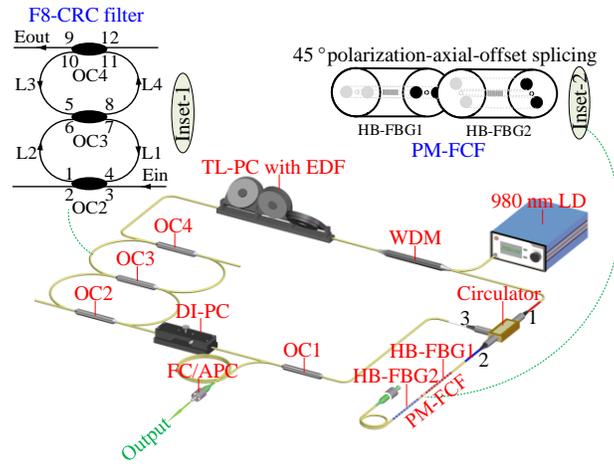

Fig. 1. Configuration of the proposed 4WS-EDFL system. The TL-PC made with a 2.9 m long EDF pigtailed by single-mode fibers (SMF-28) at both sides; Inset-1 showing the components of the F8-CRC filter, where L1~ L4 denoting lengths of fibers and 1~12 denoting port numbers of OCs; Inset-2 illustrating the 45 °polarization-axial-offset fusion splicing of PMF pigtails of HB-FBG1 and HB-FBG2, composing the PM-FCF. An FC/APC connecter used behind the PM-FCF to avoid the unnecessary reflections.

Figure 1 shows the schematic diagram of the 4WS-EDFL, simply composed of a PM-FCF, a F8-CRC filter and few other components. A ~2.9 m long EDF (Fibercore M12-980-125), coiled around the plates of a three-loop PC (TL-PC), is pumped by a 980 nm laser diode (LD, Connet VLSS-980) through a 980/1550 nm wavelength division multiplexer (WDM). The PM-FCF is innovatively used for defining the four lasing wavelengths and introducing enhanced PHB effect in the coiled EDF. A three-port circulator is used to ensure the unidirectional propagation of light. A drop-in polarization controller (DI-PC), combining with the TL-PC, can adjust light's SOP to balance the gains and losses of oscillating lasers. The F8-CRC filter, made of three OCs (OC2, OC3 and OC4) as shown in inset-1, is employed to select one expected mode from the dense longitudinal-modes of the main-ring cavity (MRC). The laser is finally extracted from the 10% port of another OC (OC1) to be measured. The length of the MRC measured by a tape-measure is ~18.25 m, determining a longitudinal-mode spacing of ~11.17 MHz.

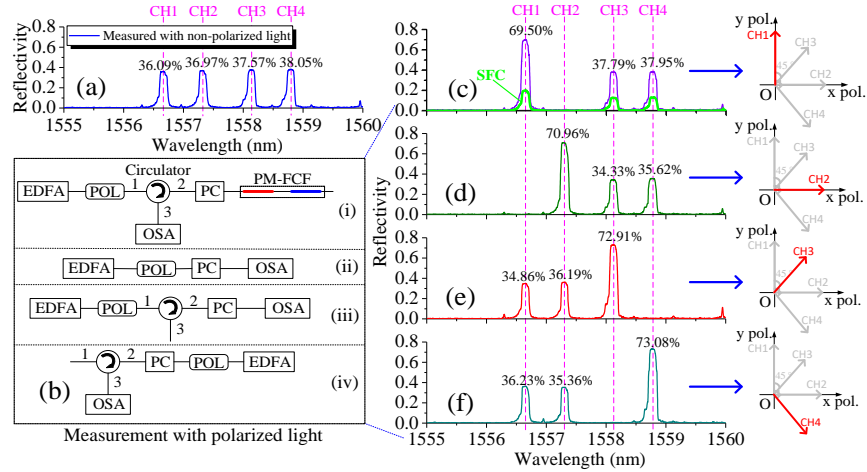

Fig. 2. (a) Pure reflection spectrum of PM-FCF in normalized linear scale measured with non-polarized light using the method described in Ref. [24]. (b) Measurement method of pure reflection spectrum of PM-FCF with polarized light, including procedures: (i) measurement of reflection spectrum of PM-FCF involving loss spectra of POL, port 1 to port 2 and port 2 to port 3 of circulator both and PC, and effect of output spectrum of EDFA; (ii) measurement of output spectrum of EDFA with the loss spectra of POL and PC; (iii) measurement of output spectrum of EDFA involving loss spectra of POL, port 1 to port 2 of circulator and PC; (iv) measurement of output spectrum of EDFA involving loss spectra of POL, PC and port 2 to port 3 of circulator. Pure reflection spectra of PM-FCF measured with polarized light, the SOP of which paralleling with the polarization-axis of (c) CH1, (d) CH2, (e) CH3 and (f) CH4, respectively. CH1, CH2, CH3 and CH4, four reflection channels of PM-FCF. SFC, spectrum for comparison in (c), measured without eliminating impact of EDFA, POL, circulator and PC.

The PM-FCF was made by fusion splicing two HB-FBGs (HB-FBG1 and HB-FBG2) together with a 45° polarization-axial-offset, as shown in the inset-2 of Fig. 1, using a specialty fiber fusion splicer (Fujikura, FSM-100P+). The HB-FBG1 and HB-FBG2 were fabricated respectively using the phase mask method [23, 24, 33, 52] in a section of hydrogen-loaded Panda-type PMF (YOFC, PM#1550_125). A uniform phase mask with a period of 1075 nm and a length of 130 mm was used. The ultraviolet (UV) light at 248 nm emitting from a KrF excimer laser was the writing light source. The lengths of two HB-FBGs were both 100 mm. Due to the different effective refractive index between the slow and fast axes of the PMF, one HB-FBG exhibits two reflection channels, corresponding to two orthogonal linear polarization modes at the two polarization-axes. Note that, in order to provide different reflection channels for HB-FBG1 and HB-FBG2, compared to writing the HB-FBG2, a suitable extra tensile-force was imposed to the PMF before writing the HB-FBG1. That was equal to using two uniform phase masks with different periods to write two HB-FBGs in turn. Therefore, in theory the PM-FCF was expected to have four reflection channels. The pure reflection optical spectrum of the PM-FCF, measured using an optical spectrum analyzer (OSA, Yokogawa AQ6370D) with a non-polarized light from a customized EDF amplifier (EDFA) on the basis of the measurement method described in [24], is shown in Fig. 2(a) in normalized linear scale. The resolution and data sampling interval of the OSA were 0.02 nm and 0.001 nm respectively.

Since the PM-FCF are polarization dependent, the reflectivities of 36.09%, 36.97%, 37.57% and 38.05% for four channels (CH1, CH2, CH3 and CH4), as marked in Fig. 2(a), measured using the non-polarized light, are not correct for a linearly polarized light input into the gratings. To measure the correct reflectivity of each channel and verify the success of 45 °polarization-axial-offset splicing between two HB-FBGs, we measured the pure reflection spectrum of the PM-FCF with a SOP controlled broadband light, based on the measurement method shown in Fig. 2(b). Four steps (i) to (iv) as described in the caption of Fig. 2 were carried out. Compared to the method in [24], an extra in-line fiber polarizer (POL) and an extra PC (the other TL-PC) were used. By adjusting the PC, a maximal reflectivity for CH1, CH2, CH3 and CH4 was measured in turn using the setup illustrated in step (i), when the polarization direction of input light was paralleling with the polarization-axis of CH1, CH2, CH3 and CH4, respectively. Then, by eliminating the impact of EDFA, POL, circulator and PC after finishing the steps (ii)–(iv), the pure reflection spectrum for each channel was obtained. The measured four pure reflection spectra are shown in Figs. 2(c)–2(f), respectively. As can be seen, the real reflectivities of four polarization-dependent channels are 69.50%, 70.96%, 72.91% and 73.08% respectively. Taking the Fig. 2(c) as an example, as expected, the reflectivity of CH2 measured is almost zero when the SOP of input light is paralleling to the polarization-axis of CH1. That is because the polarization-axes of CH1 and CH2 are orthogonal. In addition, the reflectivities of CH3 and CH4 are close to half of that of CH1, since the polarization-axes of CH3 and CH4 are respectively 45 °and 135 °orientation with respect to that of CH1, as illustrated as the coordinate system. The similar situations are seen in Figs. 2(d)–2(f), and on the basis of the results we know that the 45 ° polarization-axial-offset splicing between HB-FBG1 and HB-FBG2 is guaranteed. Note that the measured reflectivities in Figs. 2(d)–2(f) may contain some errors resulting from the limited polarization extinction ratio of the POL, inaccuracy of manually adjusting the PC and the non-negligible fusion splicing loss between the PMF and single-mode fiber (SMF). Four channels are centered at 1556.666, 1557.327, 1558.138 and 1558.795 nm, respectively, with full-widths at half maximum (FWHMs) of 0.153, 0.153, 0.139 and 0.153 nm. In addition, the first two reflection channels and the last two reflection channels are respectively with a wavelength-spacing $\Delta\lambda$ of 0.661 and 0.657 nm, corresponding to a birefringence $\Delta n$ of ~$6.14\times10^{-4}$, calculated using the equation $\Delta\lambda=\Delta n\cdot\Lambda$ ($\Lambda$ =1075 nm is the period of phase mask), which is consistent with the value for PMF in 1550 nm band. Note that the birefringence could be changed to some extent by the high-pressure hydrogen-loading process compared to that measured by the manufacturer before the PMF leaves the factory.

*2.2 Study on F8-CRC filter for SLM operation*

The F8-CRC filter, as shown in Fig. 1, is assembled by three 2×2 fiber OCs (OC2, OC3 and OC4) with two sub-rings. Here, we introduce a graphical approach, called the signal-flow graph method, to analyze the optical fiber based F8-CRC filter. The signal-flow graph was originally proposed by Mason [53, 54], to obtain the causal relationship of the signal transformation and transmission in electrical circuits. Compared to the conventional methods [23, 24], this method is exceedingly helpful for understanding the CRC filter system operation using a pictorial representation. It also allows one to easily and systematically manipulate the variables of interest for precisely designing and studying a novel CRC filter. In addition, one can directly obtain the filtering-system's transfer function using the Mason' rule [53, 54], without taking complicated mathematical operation. We set 1–12 and $E1$–$E12$ as the light ports of the OCs and corresponding electric-field amplitudes respectively, $\kappa_i(i=2,3,4)$ and $\gamma_i(i=2,3,4)$ as the cross-coupling ratio and insertion loss of the $i$th OC respectively, $L1$–$L4$ as the fiber lengths, $\alpha$ as the fiber loss coefficient, $\delta$ as the fusion splicing loss, and $\beta=2\pi n_{eff}/\lambda$ as the light propagation constant. Where, $n_{eff}$ is the effective refractive index and $\lambda$ is the wavelength. The F8-CRC filter can be pictorially represented as Fig. 3. 12 light ports are represented as photonics nodes and 11 of them are involved. In order to analyze the transitive relation from node 3 ($E_{in}$: $E_3$) to node 9 ($E_{out}$: $E_9$) using Mason's rule expressed as Eq. 1, we need to analyze the dependence of 11 nodes among each other and the path and loop gains of signal-flows firstly, based on the signal-flow graph in Fig. 3.

$$H = \frac{1}{\Delta} \sum_i P_i \Delta_i , \quad (1)$$

where, $H$ is the network function relating an input and an output port, $P_i$ is the gain of the $i$th forward optical path from $E_{in}$ to $E_{out}$. The $\Delta$ denotes the determinant of the signal-flow graph, given as

$$\Delta = 1 - \sum_i Q_i + \sum_{l \neq m} Q_l Q_m - \sum_{l \neq m \neq n} Q_l Q_m Q_n + \cdots , \quad (2)$$

where $\sum_i Q_i$ is the sum of all loop gains, $\sum_{l \neq m} Q_l Q_m$ is the sum of products of all loop gains of two non-touching optical loops, and $\sum_{l \neq m \neq n} Q_l Q_m Q_n$ is the sum of products of all loop gains of three non-touching optical loops. $\Delta_i$ in Eq. (1) denotes the determinant $\Delta$ after all loops that touch the path $P_i$ at any node have been eliminated.

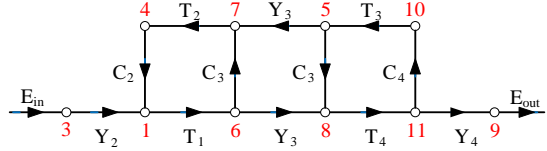

Fig. 3. Signal-flow graph representation of the F8-CRC filter.

As shown in Fig. 3, we define the straight transmittance $C_i$ of three OCs as

$$C_i = \sqrt{1-\gamma_i}\sqrt{1-\kappa_i} \quad (i = 2,3,4), \quad (3)$$

the cross-coupling transmittance $Y_i$ of three OCs as

$$Y_i = i\sqrt{\kappa_i}\sqrt{1-\gamma_i} \quad (i = 2,3,4), \quad (4)$$

and the propagating gain $T_i$ of fiber optical paths $L_i$ as

$$\begin{cases} T_1 = \sqrt{1-\delta}e^{(-\alpha+j\beta)L_1} \\ T_2 = \sqrt{1-\delta}e^{(-\alpha+j\beta)L_2} \\ T_3 = \sqrt{1-\delta}e^{(-\alpha+j\beta)L_3} \\ T_4 = \sqrt{1-\delta}e^{(-\alpha+j\beta)L_4} \end{cases} . \quad (5)$$

There are 3 loops in Fig. 3, respectively involving the optical nodes 1–6–7–4–1, 8–11–10–5–8, and 1–6–8–11–10–5–7–4–1, and the propagating gain $Q_i$ can be given by

$$\begin{cases} Q_1 = T_1 C_3 T_2 C_2 \\ Q_2 = T_4 C_4 T_3 C_3 \\ Q_3 = T_1 Y_3 T_4 C_4 T_3 Y_3 T_2 C_2 \end{cases} . \quad (6)$$

As can be seen, $Q_1$ and $Q_2$ are from two non-touching optical loops. So, according to Eq. (2),

the determinant $\Delta$ can be expressed as

$$\Delta = 1 - (Q_1 + Q_2 + Q_3) + Q_1 Q_2. \tag{7}$$

Only one forward optical path from $E_{in}$ to $E_{out}$, involving the nodes 3–1–6–8–11–9, can be found in Fig. 3, and the propagating gain $P_1$ of which can be given as

$$P_1 = Y_2 T_1 Y_3 T_4 Y_4. \tag{8}$$

According to Eq. (2), the corresponding $\Delta_i$ can be known as

$$\Delta_i = 1. \tag{9}$$

Therefore, based on the Mason's rule, the transmission function $H$ from $E_{in}$ to $E_{out}$ can be obtained as

$$H = \frac{E_9}{E_3} = \frac{P_1 \Delta_1}{\Delta}. \tag{10}$$

Subsequently, the transmittance $T$ of the F8-CRC filter can be derived from Eqs. (3)–(10), and is expressed as

$$T = \frac{E_9}{E_3} \cdot \left(\frac{E_9}{E_3}\right)^*$$
$$= \frac{\kappa_2 \kappa_3 \kappa_4 (1-\gamma_2)(1-\gamma_3)(1-\gamma_4)(1-\delta)^2 e^{-2\alpha(L_1+L_4)}}{1+x^2+y^2+z^2-2(x+yz)\cos[\beta(L_1+L_2)]-2(y+xz)\cos[\beta(L_3+L_4)]+2xy\cos[\beta(L_1+L_2-L_3-L_4)]+2z\cos[\beta L]} \tag{11}$$

where the parameters $x$, $y$ and $z$ are given as

$$\begin{cases} x = \sqrt{1-\gamma_2}\sqrt{1-\gamma_3}\sqrt{1-\kappa_2}\sqrt{1-\kappa_3}(1-\delta)e^{-\alpha(L_1+L_2)} \\ y = \sqrt{1-\gamma_3}\sqrt{1-\gamma_4}\sqrt{1-\kappa_3}\sqrt{1-\kappa_4}(1-\delta)e^{-\alpha(L_3+L_4)} \\ z = \sqrt{1-\gamma_2}\sqrt{1-\gamma_4}\sqrt{1-\kappa_2}\sqrt{1-\kappa_4}(1-\gamma_3)(1-\delta)^2 e^{-\alpha L} \\ L = L_1 + L_2 + L_3 + L_4 \end{cases} \tag{12}$$

Based on the theory above, we can simulate the transmission spectrum of the F8-CRC filter. Combining with the measured reflection spectrum of PM-FCF, we can design the fabrication parameters of the F8-CRC filter aiming for selecting one longitudinal-mode from dense modes in the MRC of the proposed EDFL. According to our previous study [24], for a ring fiber laser with a FBG as the initial lasing wavelength decider and an ultra-narrow comb filter as the SLM selector, the FSR and passband of the comb filter must be larger than 0.5 times of the FBG's reflection bandwidth and less than 2 times of the MRC's mode-spacing, respectively. The FWHM for every reflection peak of the PM-FCF is ~0.15 nm, corresponding to a frequency range of ~18 GHz, and the mode-spacing of the MRC is calculated to be ~11.17 MHz. By comprehensively considering the theory, transmittance, feasibility and handwork tolerance of fabrication for the F8-CRC filter and after optimizing the simulations, we finally used the parameters as $L_1 = L_3 = L_4 = 0.25$ m, $L_2 = 0.27$ m, $\kappa_i (i=2,3,4) = 0.05$, $\alpha = 0.2$ dB/km, $\delta = 0.01$ dB, $\gamma_i (i=2,3,4) = 0.09$ dB and $n_{eff} = 1.468$. The simulated transmission spectrum of the F8-CRC filter is shown as the blue solid-line in Fig. 4(a-1), and the measured reflection spectrum of the PM-FCF is also plotted using the red dashed-line. As can be seen, the FSR of the F8-CRC filter is 10.20 GHz and only one transmission peak is located in the FWHM of every channel of the PM-FCF. Multiplying the two curves in Fig. 4(a-1) by each other, we obtain the combined filtering effect of the F8-CRC filter and PM-FCF, as shown in Fig. 4(b-1). Four narrow channels are seen and expected to be utilized to achieve wavelength-switchable lasing in the 4WS-EDFL. By zooming in to see the maximum transmission peak in the third channel as shown in the inset-1 of Fig. 4(b-1), there are two passbands both with a 3-dB bandwidth as narrow as ~8.75 MHz, which allows only one longitudinal-mode to pass through.

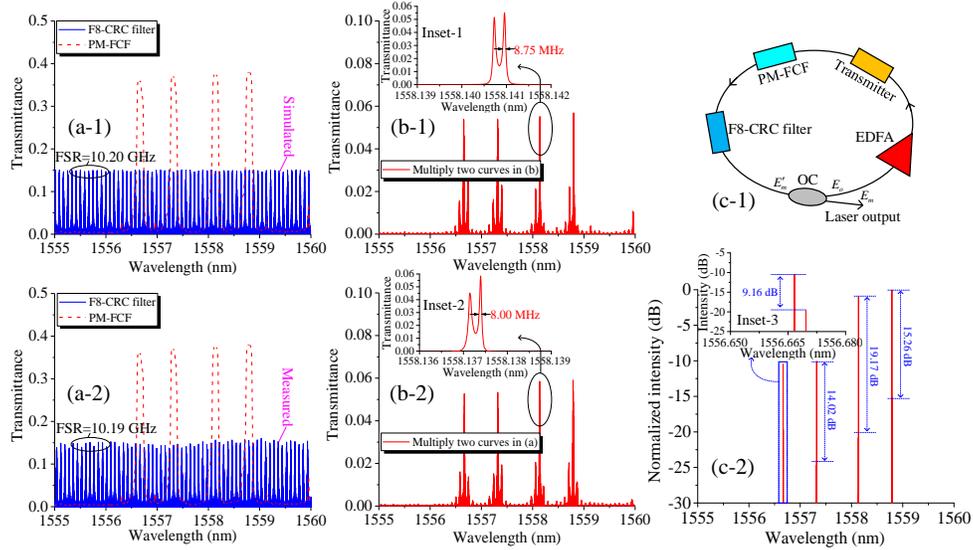

Fig. 4. (a-1) Simulated transmission spectrum of F8-CRC filter plotted by blue solid-line, with measured reflection spectrum of PM-FCF plotted by red dashed-line for comparison; (b-1) Combined filtering effect of F8-CRC filter and PM-FCF obtained by multiplying two curves in (a-1); Inset-1 showing the enlarged maximum transmission peak in the third channel in (b-1). (a-2) Measured transmission spectrum of F8-CRC filter plotted by blue solid-line, with measured reflection spectrum of PM-FCF plotted by red dashed-line for comparison; (b-2) Combined filtering effect of F8-CRC filter and PM-FCF obtained by multiplying two curves in (a-2); Inset-2 showing the enlarged maximum transmission peak in the third channel in (b-2). (c-1) Simple ring EDFL model for verifying the capability of F8-CRC filter combining with the PM-FCF to select SLM theoretically; (c-2) Simulated laser oscillating for 20 loops for the model in (c-1) but without considering the gain saturation effect and mode competition in EDF induced by homogeneous broadening effect; Inset-3 showing the enlarged laser oscillating in the first channel in (c-2).

Using the simulation parameters, we fabricated the F8-CRC filter using three commercial OCs all with a coupling ratio of 95:5, trying our best to minimize the handwork error. Its transmission spectrum was measured under room temperature, by inputting a sweeping-laser from a tunable laser source (Yenista T100S-HP) into the filter and detecting its output using a 400 MHz photodetector (PD, Thorlabs PDB470C). The photovoltage of the PD was monitored by a data acquisition card (DAQ, Measurement Computing Cor. USB-1602HS). The laser source was with a sweeping-speed of 1 nm/s and a sweeping-range of 1554–1561 nm, and the DAQ was with a data sampling rate of 500 kHz. In the data processing, the laser wavelength vs. data sampling interval and the laser power vs. PD's photovoltage were calibrated each other, respectively. Figure 4(a-2) plots the measured result using the blue solid-line and also plots the reflection spectrum of the PM-FCF using the red dashed-line. As can be seen, the experimental result in Fig. 4(a-2) is in good agreement with the corresponding simulated spectrum of the F8-CRC filter in Fig. 4(a-1), and the measured FSR of 10.19 GHz is very close to the simulated value of 10.20 GHz. Figure 4(b-2) gives the combined filtering effect based on the measured curves in Fig. 4(a-1), using the same data processing method to obtain Fig. 4(b-1). And also, high consistency is seen between the simulated and measured results respectively shown in Fig.

4(b-1) and Fig. 4(b-2). The measured passband of 8.00 MHz shown in the inset-2 of Fig. 4(b-2) is close to that of 8.75 MHz marked in the inset-1 of Fig. 4(b-1). The high consistency between the simulations and experiments validates that our research approach for CRC filter design is with high feasibility and accuracy. The little inconsistency is mainly from the parameter errors of coupling ratios and insertion losses of commercial OCs, fusion splicing losses and fiber lengths in the filter fabrication with them used in the simulations.

To verify the capability of our F8-CRC filter combining with the PM-FCF to select SLM in a fiber laser theoretically, we build a simple ring EDFL model as shown in Fig. 4(c-1). In that, an EDFA is introduced as the gain medium, the F8-CRC and the PM-FCF are both abstracted to transmission filters, an OC with a coupling ratio of 90:10 outputs the laser from the 10% port, a transmitter represents the total loss including the extra insertion losses of above components and the passive fiber loss. We define the transmission functions of F8-CRC and PM-FCF as $T_{CRC}$ and $T_{FBG}$ respectively, and subsequently the combined transmittance given in Fig. 4(b-2) is exact $T_{CRC}T_{FBG}$. Here, we introduce a small signal gain $G$ for the EDFA, and assume that the loss of the transmitter is $A$. Then, for a small signal with an amplitude of $E_o$ oscillating inside the laser cavity for $m$ loops, the laser amplitude $E'_m$ can be given as

$$E'_m = \frac{\left[E_o\sqrt{G}\sqrt{A}\sqrt{T_{CRC}T_{FBG}}\,\mathrm{e}^{(-\alpha L+j\beta L_c)}\right]\left(1-q^m\right)}{1-q} \ . \tag{13}$$

Where $L_c$ denotes the laser cavity length, and the factor $q$ is given as

$$q = \sqrt{G}\sqrt{A}\sqrt{T_{CRC}T_{FBG}}\sqrt{1-\kappa}\,\mathrm{e}^{(-\alpha L+j\beta L_c)} \ . \tag{14}$$

Then the output amplitude from the OC can be obtained as

$$E_m = \sqrt{\kappa}\cdot E'_m = \sqrt{\kappa}\cdot\frac{\left[E_o\sqrt{G}\sqrt{A}\sqrt{T_{CRC}T_{FBG}}\,\mathrm{e}^{(-\alpha L+j\beta L_c)}\right]\left(1-q^m\right)}{1-q}, \tag{15}$$

where $\kappa$ is the cross-coupling ratio of the OC. Finally, the simulated laser output from the cavity can be derived from Eq. (13)–(15) as

$$I_{out} = E_m \cdot E_m^* = \frac{E_o^2 q^2 \kappa\left[1-2\left(qe^{-j\beta L_c}\right)^m \cos(\beta L_c m)+\left(qe^{-j\beta L_c}\right)^{2m}\right]}{(1-\kappa)\left[e^{2j\beta L_c}-2q\cos(\beta L)e^{j\beta L_c}+q^2\right]} \ . \tag{16}$$

Note that, in the modeling, we ignore both the gain saturation effect and homogeneous broadening induced mode competition presented in the practical EDFA. It is reasonable that the laser output intensity will be infinity for an infinite number of oscillating loops, as can be inferred from Eq. (15).

On the basis of the laser model above, we assume $G = 20$ dB, $A = -5$ dB, $L_c = 18.257$ m same with that of the experimental laser cavity, and $\kappa = 0.1$. For a typical number of oscillating loops $m = 20$ the laser output can be calculated as shown in Fig. 4(c-2). As can be seen, four lasing lines with side-mode suppression ratios (SMSRs) of 9.16 (as marked in inset-3), 14.02, 19.17 and 15.26 dB respectively are obtained, corresponding to the four channels of the PM-FCF. Note that the intensity labeled in the y-axis of Fig. 4(c-2) is normalized to one using the maximum of the four main peaks and is converted into decibel. Considering that in the EDF there is strong homogeneous broadening induced mode competition, a SMSR of >9.16 dB can definitely enable the SLM lasing in each channel of the PM-FCF theoretically. Therefore, the fabricated F8-CRC filter combining with the PM-FCF is able to achieve SLM operation for the proposed 4WS-EDFL in all four channels in theory.

*2.3 Principle of wavelength-switchable operation*

The EDF is a typical homogenous broadening gain medium, so under room temperature severe mode/wavelength competition is presented inside the laser cavity of an EDFL. In addition, each channel's reflectivity of the PM-FCF is strong polarization-dependent, so the loss of an oscillating laser at any channel can be easily controlled by adjusting the DI-PC and the TL-PC carefully. That is the major part of PDL inside the laser cavity. On the basis of above two aspects, one can make the lasing at an expected wavelength suffer lowest loss, and consequently the proposed 4WS-EDFL can easily achieve single-wavelength switching-operation among four wavelengths. However, one may not achieve stable multi-wavelength lasing through only controlling the PDL to suppress the severe wavelength competition in an EDFL. In the proposed 4WS-EDFL, we introduced the enhanced PHB effect for achieving multi-wavelength operation. The PHB in an EDF arises from the randomly distributed orientations of erbium ions in the glass matrix and the selective deexcitation of those ions aroused by a polarized light [55]. The lights reflected by PM-FCF with different SOPs can utilize different subsets of excited erbium ions, indicating that the gains of the lights with different SOPs are contributed by different groups of ions. That can mitigate wavelength competition significantly. As aforementioned, the PM-FCF is made of two HB-FBGs fusion spliced together with a 45 °polarization-axial-offset and can reflect four wavelengths all with different SOPs. So they can be amplified by the different excited erbium ions. In addition, the gain EDF coiled around the three circular plates of the TL-PC can introduce slight stress inside itself and therefore a birefringence depending on the radius of the EDF and the radius of the plates [44]. In the coiled EDF, the orientations of the polarization axes depend on the angular orientations of the three plates, meaning that a rotation of the plates will rotate the two polarization axes. Therefore, by tuning the TL-PC, the polarization axes of the coiled EDF can be aligned with the polarization directions of the incident lights at different wavelengths. Also, the output SOPs from the TL-PC can be controlled to well match with the expected channels' SOPs of the PM-FCF. Then the net gains at the expected reflection wavelengths of the PM-FCF are almost equal, which further enhances the PHB effect inside the laser cavity. Finally, the expected wavelengths experience same small-signal gain and achieve lasing simultaneously with high stability.

## 3. Experimental results and discussion

The proposed 4WS-EDFL system was constructed on an ordinary steel optical table, using soft adhesive tapes to fix loose fibers and a foam box to cover the F8-CRC filter. All experiments were carried out under laboratory room temperature (with air conditioning constantly running), with as much as possible quiet environment. When the pump power of 980 nm LD was beyond the threshold, we achieved fifteen lasing states, including four single-, six dual-, four tri- and one quad-wavelength operations through adjusting the TL-PC and DI-PC carefully to tune the polarization-axes of the coiled EDF and the SOP of light inside the laser cavity. We emphasized to study the lasing characteristics of single- and dual-wavelength operations and the photonic generation of microwave signal with the dual-wavelength lasing output of the 4WS-EDFL.

*3.1 Single-wavelength operation*

Using a pump power of 200 mW for demonstration, we obtained stable single-wavelength operations for the 4WS-EDFL lasing at $\lambda 1$, $\lambda 2$, $\lambda 3$ and $\lambda 4$ respectively, as shown in Figs. 5(a)–5(d). In each graph, 15 repeated optical spectrum scans were measured in a time span of ~60 min, by the AQ6370D OSA using a resolution of 0.02 nm and a data sampling interval of 0.001 nm, to demonstrate the lasing stability. As can be seen in all graphs, we give the key parameters extracted from the three-dimensional (3-D) curves. The four lasers are with little to no wavelength fluctuation $f_{\lambda i}$ (i=1, 2, 3, 4) (Maximum: 0.008 nm, less than the resolution of OSA) and low power fluctuation $f_{pi}$ (Maximum: 0.692 dB) lasing at $\lambda i$ (i=1, 2, 3, 4). Note that, since the OSA's resolution is limited, a more accurate wavelength fluctuation may be measured by a wavemeter with a higher resolution. The lasers respectively concentrated at ~1556.45 nm ($\lambda 1$), ~1557.12 nm ($\lambda 2$), ~1557.98 nm ($\lambda 3$) and ~1558.65 nm ($\lambda 4$) are basically consistency with the center wavelengths of the PM-FCF' four channels. We believe that the little wavelength deviation and slight instability aforementioned were mainly induced by the additional force from the fixation of PM-FCF, the fluctuation of ambient temperature, the RIN of pump LD and the inevitable mechanical vibrations of surroundings. In addition, the OSNRs are all higher than 82 dB and the SMSRs are all larger than 67 dB for the four lasers, indicating that the laser cavity design is excellent with a high quality-factor.

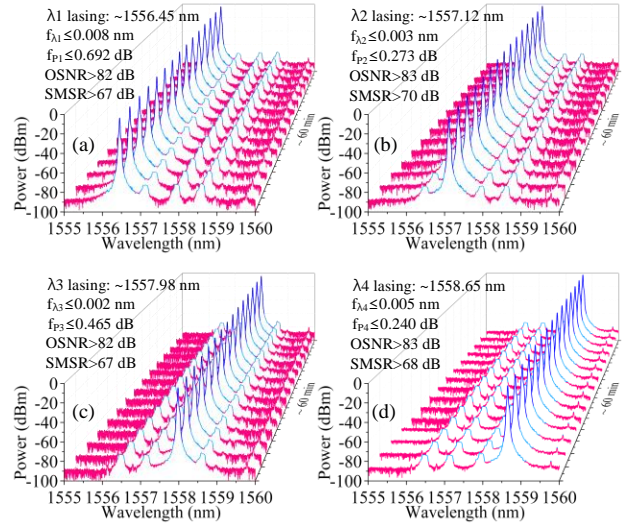

Fig. 5. Spectra of single-wavelength switchable operations lasing at (a) $\lambda 1$, (b) $\lambda 2$, (c) $\lambda 3$ and (d) $\lambda 4$, respectively, measured in a time span of ~60 min. $f_{\lambda i}$ (i=1, 2, 3, 4): fluctuation of wavelength lasing at $\lambda i$; $f_{pi}$ (i=1, 2, 3, 4): fluctuation of power lasing at $\lambda i$; In each figure, 15 repeated spectra measured by OSA with a time interval of ~4 min.

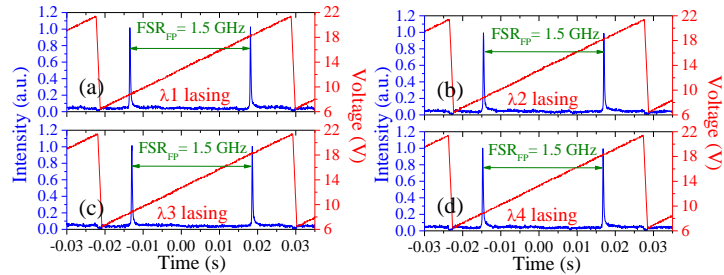

Fig. 6. Longitudinal-mode characteristics, measured by a scanning Fabry-Pérot interferometer, in single-wavelength operations lasing at (a) $\lambda 1$, (b) $\lambda 2$, (c) $\lambda 3$ and (d) $\lambda 4$, respectively.

The longitudinal-mode characteristics of the 4WS-EDFL at each lasing wavelength were firstly investigated by a scanning Fabry-Pérot (F-P) interferometer (Thorlabs, SA200-12B) with a FSR of 1.5 GHz and a resolution of 7.5 MHz. As can be seen in Figs. 6(a)–6(d), there is only one peak captured in a FSR of 1.5 GHz for each laser, indicating that the 4WS-EDFL was operating in a stable SLM state. To further confirm the SLM operation, we measured each laser output in turn, using the self-homodyne method with a 400 MHz PD and a radio frequency (RF) electrical spectrum analyzer (ESA, Keysight N9010A), as shown in Fig. 7(a). The parameter setup of the ESA can be found in the caption of Fig. 7. As expected, in a ~10 min measurement there is no any beating signal captured for every laser. In order to investigate the mode selection capability of the F8-CRC filter, we replaced it by a section of SMF to maintain the original MRC's length and then measured each laser output again, as shown in Fig. 7(b). Numerous spikes are seen, indicating that every laser was with dense longitudinal-modes. In addition, the minimum spacing of two adjacent peaks is ~11.20 MHz, which is high consistent with the calculated longitudinal-mode spacing of 11.17 MHz of the MRC. Note that, since the passband of F8-CRC filter, bandwidth of PM-FCF and mode-spacing of MRC are all pump power independent, the SLM operation can be guaranteed under different pump powers.

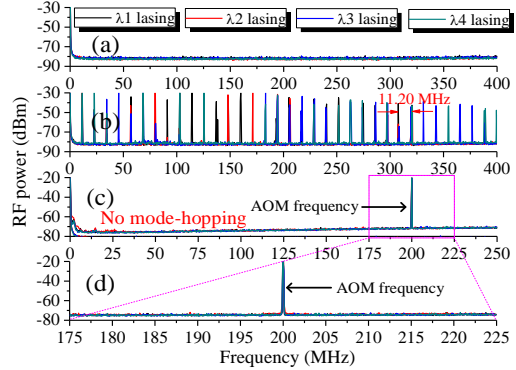

Fig. 7. RF spectra measured by ESA for all single-wavelength operations in turn, plotted by different colored curves. Self-homodyne RF spectra measured in a range of 0–400 MHz using maximum-hold (MH) mode in ~10 min with resolution bandwidth (RBW) of 51 kHz, (a) when F8-CRC filter connected and (b) when F8-CRC filter replaced by SMF with a same length. Delayed self-heterodyne RF spectra measured using MH mode in ~30 min, in ranges of (c) 0–250 MHz with RBW of 51 kHz and (d) 175–225 MHz with RBW of 30 kHz.

Same with that described in previous publications [23, 24], we studied every laser's mode-hop characteristic using a delayed self-heterodyne measurement system (DSHMS) composed of a 400 MHz PD, a Mach-Zehnder interferometer (MZI) with a 200 MHz acoustic optical modulator (AOM) and 100 km long SMF in two arms respectively, and the RF ESA. In a ~30 min measurement, the results in ranges of 0–250 MHz and 175–225 MHz are shown in Figs. 7(c) and 7(d) respectively. Due to the long delay-line in MZI and the longitudinal-mode spacing of ~11.20 MHz, any mode-hopping should be captured, but only the strong beating signal at ~200 MHz introduced by the AOM was captured for every laser. Therefore, we believe that our 4WS-EDFL has the potential to work in a stable SLM operation at any one of the four wavelengths without mode-hop for a long time. Note that, the mode-hopping may occur and be captured during the wavelength switching process.

The linewidths of four lasers were also measured using the DSHMS as shown in Figs. 8(a)–8(d), with the parameter setup of ESA given in the caption of Fig. 8. The measured RF beating spectra of λ1, λ2, λ3 and λ4 lasings are all curve-fitted well using the Lorentz lineshape, respectively with a high adjusted $R$-Square (Adj. $R$-Square) of 0.9933, 0.9926, 0.9924 and 0.9927. As marked and calculated in four graphs, the linewidths are 584, 590, 592 and 598 Hz respectively. It is worth noting that, since we could not use a SMF delay-line over 1500 km to achieve completely incoherent mixing of two arms of the MZI due to the limitation of output laser power and serious $1/f$ frequency noise from the ultra-long delay-line, it is impossible to obtain a pure Lorentz linewidth spectrum. However, considering that the measured results contain unavoidable broadening effects induced by the partial coherence mixing of the MZI's two arms and $1/f$ frequency noise from the 100 km delay-line, we believe the lasers' linewidths obtained must be larger than their real values. Therefore, the measured values can be regarded as conservative characterizations of the natural linewidths for future practical applications.

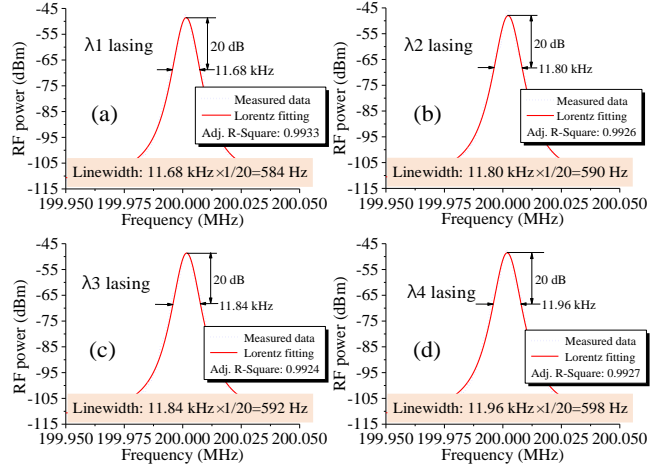

Fig. 8. Measurements of laser linewidths for all single-wavelength operations in turn using the DSHMS. Delayed self-heterodyne RF beating spectra of laser outputs oscillating at (a) λ1, (b) λ2, (c) λ3 and (d) λ4, respectively, in 199.950–200.050 MHz using average mode (100 times) of ESA with RBW of 100 Hz. In each figure, measured data fitted by Lorentz lineshape.

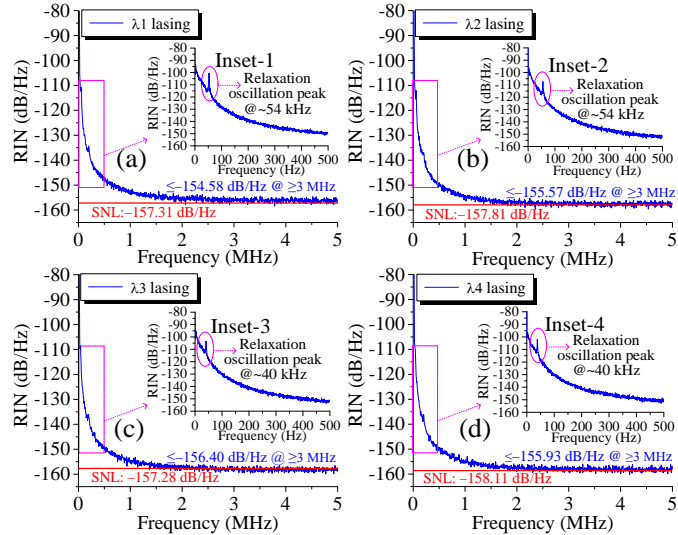

Fig. 9. RIN spectra of single-wavelength operations lasing at (a) λ1, (b) λ2, (c) λ3 and (d) λ4, respectively, in 0–5 MHz using RBW of 10 kHz; for comparison, in each figure, corresponding shot-noise limit (SNL) also given and plotted; insets showing the same measurements in 0–500 kHz using RBW of 100 Hz for ESA to display the relaxation oscillation peaks.

The RIN parameter is generally used to characterize the instantaneous power stability of a SLM laser. We measured the RIN spectra of four lasers in turn using the 400 MHz PD, an oscilloscope (Tektronix, TDS2024C) and the ESA, as shown in Figs. 9(a)−9(d), using the parameter setup of ESA as given in the caption of Fig. 9. For comparison, in each graph, the shot noise limit (SNL) for each laser is given, which can be calculated as

$$\text{SNL} = 10\lg\left(\frac{2h\nu}{P}\right) \text{ (dB/Hz)}, \tag{16}$$

where, $h$ is the Planck constant, $\nu$ is the laser frequency and $P$ is the laser output power (the measured values for lasing wavelengths λ1, λ2, λ3 and λ4 are ~1.38, ~1.55, ~1.37 and ~1.66 mW, respectively). As can be seen, when the frequency is larger than 3 MHz, the RIN of our fiber laser is ≤−154.58 dB/Hz for all four output lasers, and for each laser the RIN@≥3 MHz is close to the corresponding SNL. In addition, we measured the relaxation oscillation peaks for four lasers as shown in the insets. The laser's relaxation oscillation noise is mainly induced by the fluctuations of pump power and cavity loss, mechanical vibration, and thermal disturbance [56]. As can be seen, all peak values are ≤−100 dB/Hz. The above data indicates that our 4WS-EDFL has good instantaneous power stability. Besides, the relaxation oscillation frequencies of lasers λ1 and λ2 are both ~54 kHz and lasers λ3 and λ4 are both ~40 kHz respectively. Theoretically, a longer MRC length determines a lower relaxation oscillation frequency [14]. For our 4WS-EDFL, the lasers λ3 and λ4 have longer cavity length than that of lasers λ1 and λ2, due to the fiber separation between two HB-FBGs. Note that all of the relaxation oscillation frequencies are lower than that of ~60 kHz measured in our former work [23] since in this work the MRC length is longer, which is also consistent with the theory [14].

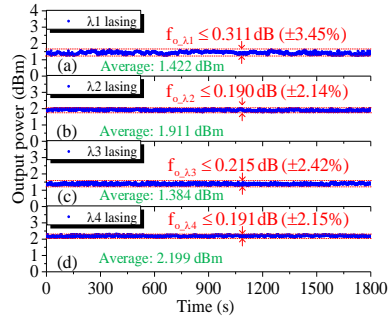

Fig. 10. Output power stabilities lasing at (a) λ1, (b) λ2, (c) λ3 and (d) λ4, respectively, measured by a laser power meter using a data sampling rate of 1 Hz in a time span of 30 min. $f_{o\_\lambda i}$ (i=1, 2, 3, 4): fluctuation of output power lasing at λi. The average output power for each laser also given.

The medium-term output power stability of the 4WS-EDFL was studied by measuring the four lasers' powers respectively in a 30 min time span using a power meter with a data sampling rate of 1 Hz, as shown in Figs. 10(a)–10(d). As can be seen, the average powers are 1.422 dBm (1.387 mW), 1.911 dBm (1.553 mW), 1.384 dBm (1.375 mW) and 2.199 dBm (1.659 mW). The slight discrepancy among each other is mainly induced by the reflectivity difference of four channels of PM-FCF and the difference of net PDL losses inside the laser cavity due to the adjusting of PCs for four lasers. Furthermore, the power fluctuations are marked in the graphs. The maximum is as low as 0.311 dB (±3.45%) at λ1 lasing. The little fluctuation is mainly due to the slight mismatch between the two subrings of the F8-CRC filter or between the F8-CRC filter and the PM-FCF induced by the temperature fluctuations and mechanical vibrations.

Polarization control is a special feature in our 4WS-EDFL. Since the fiber laser cavity is not all polarization maintaining, the SOP of laser output from the pigtailed SMF jumper of OC1 cannot be guaranteed to be linear polarization, although the light reflected from any channel of the PM-FCF is linearly polarized. In order to study the SOP characteristics of laser outputs, we measured the SOPs of four lasers in turn using a polarization analyzer (General Photonics Cor., PSY-201) as shown in Figs. 11(a)–11(d). Each measurement was accumulated continuously for 5 min. To avoid the laser's SOP being influenced by the instantaneous external disturbance, the output SMF jumper connecting to the PSY-201 was fixed very well using soft adhesive tapes during the measurements. As can be seen, in every image the SOP data trace on the surface of Poincaré sphere is concentrated in a small region and the degree of polarization (DOP) is close to 100% (here, the DOP values slightly larger than 100% for λ1 and λ3 lasing are induced by the analysis error of equipment). That indicates that the four lasers' SOPs are all with high stability and excellent single-polarization property. Additionally, from the polarization ellipse on the top-right corner of each image, we see that the ellipses' major axes of lasers λ1 and λ2/lasers λ3 and λ4 are orthogonal respectively while the ellipses' major axes of lasers λ1 and λ3/lasers λ2 and λ4 are 45° with respect to each other, consistent with that analyzed from Fig. 2 in Section 2A.

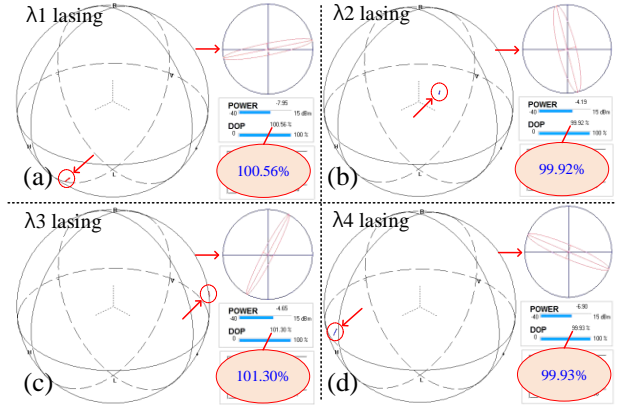

Fig. 11. SOP measurements of single-wavelength operations lasing at (a) λ1, (b) λ2, (c) λ3 and (d) λ4, respectively.

The above performance characterizations for four lasers in single-wavelength operation of our 4WS-EDFL validate that our laser system design is superior. We also believe that its output performance can be greatly improved further if a specialized packaging of vibration isolation and temperature control is employed in future for practical applications. In addition, since the number of lasing wavelengths is mainly decided by the PM-FCF, we believe that more lasing wavelengths may be obtained if a new filter with more channels combining with specific polarization-management can be proposed and used in our current laser configuration.

*3.2 Multi-wavelength operation*

Based on the mechanism of 45 °polarization-axial-offset fusion splicing between the HB-FBG1 and HB-FBG2 and the analysis on the polarization directions of lights reflected from different channels of the PM-FCF in Section 2A, we know that there is no any two lasing wavelengths among λ1, λ2, λ3 and λ4 of the 4WS-EDFL possessing same SOP. Actually, the SOPs' major ellipse axes of any two lasers are 45 °or an integer multiple of 45 °with respect to each other. That can enable dual-wavelength lasing stably on the basis of the enhanced PHB introduced in the EDF coiled in the TL-PC. Still using the 200 mW pump power for demonstration, through adjusting the two PCs carefully, we obtained the dual-wavelength lasing at λ1&λ2, as shown as the 3-D spectrum in Fig. 12 (a). Same with the measurement methods used for single-wavelength operation, the self-homodyne beating spectrum for demonstrating SLM lasing, the delayed self-heterodyne spectrum for demonstrating mode-hop free, the RIN spectrum for demonstrating high instantaneous power stability were measured as shown in Figs. 12(b)–(d) respectively. The parameter setup of instruments is given in the caption of Fig. 12. Note that since the spacing of λ1 and λ2 is ~0.67 nm, corresponding to a frequency spacing of ~80.4 GHz, the beating signal between λ1 and λ2 lasing cannot be detected by the 400 MHz PD used in the self-homodyne and self-heterodyne systems. As can be seen in Fig. 12, the dual-wavelength operation is with high optical spectrum and longitudinal-mode stability. The key property parameters are shown in Table 1. Similarly, we also obtained other five groups of dual-wavelength lasing combinations, including λ1&λ3, λ1&λ4, λ2&λ3, λ2&λ4, λ3&λ4, all with excellent performances. For saving space we will not show more figures similar with Fig. 12, but list all of the key property parameters in Table 1. As can be seen, all of the six dual-wavelength operations possess extremely low wavelength and power fluctuations, high OSNRs and SMSRs, and low RINs. Especially, six dual-wavelength operations are with very similar performance, which is consistent with the expectation, benefiting from the specially enhanced PHB effect. Moreover, all dual-wavelength operations are with the performance parameters close to those of the single-wavelength operations. The linewidth for each lasing wavelength should be measured through selecting it from two simultaneously lasing ones, using an ultra-narrowband tunable optical filter with ultra-high stopband suppression. However, we did not have such a filter, so we only directly observed the self-heterodyne beating spectra of the laser outputs of six dual-wavelength operations respectively, and the conservative linewidths mixing two lasers were all less than 800 Hz. In addition, in order to further prove the good performance of our 4WS-EDFL, we compares the key stability parameters of dual-wavelength operations achieved with some typical techniques, as listed in Table 2. The data indicates that the proposed 4WS-EDFL has an outstanding medium-term stability, which enables it to be a good switchable dual-wavelength laser source for many related important applications, especially after it is packaged by specialized vibration isolation and temperature control techniques.

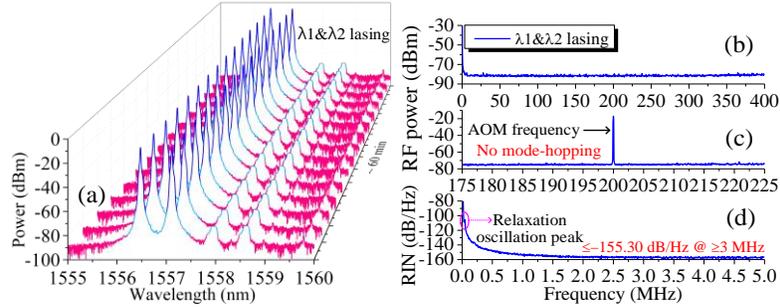

Fig. 12. Measurements of optical spectra and RF spectra in dual-wavelength operation lasing at $\lambda_1\&\lambda_2$. (a) 15 repeated OSA scans with a time interval of ~4 min; (b) Self-homodyne RF spectrum measured in a range of 0–400 MHz using maximum-hold (MH) mode in ~10 min with resolution bandwidth (RBW) of 51 kHz; (c) Delayed self-heterodyne RF spectrum measured in a range of 0–250 MHz using MH mode in ~30 min with RBW of 51 kHz; (d) RIN spectrum measured in 0–5 MHz using RBW of 10 kHz.

**Table 1. Performance parameters of dual-wavelength lasing output**

| Lasers[a] | Wavelength Fluctuation (nm) | | Power Fluctuation (dB) | | OSNR (dB) | SMSR (dB) | RIN (dB/Hz) | Linewidth (Hz) |
|---|---|---|---|---|---|---|---|---|
| $\lambda_1 \& \lambda_2$ | $f_{\lambda_1}\leq 0.005$ | $f_{\lambda_2}\leq 0.006$ | $f_{P_1}\leq 0.851$ | $f_{P_2}\leq 0.952$ | 79 | 65 | −155.30 | |
| $\lambda_1 \& \lambda_3$ | $f_{\lambda_1}\leq 0.014$ | $f_{\lambda_3}\leq 0.008$ | $f_{P_1}\leq 0.995$ | $f_{P_3}\leq 0.971$ | 78 | 61 | −154.94 | |
| $\lambda_1 \& \lambda_4$ | $f_{\lambda_1}\leq 0.006$ | $f_{\lambda_4}\leq 0.004$ | $f_{P_1}\leq 0.978$ | $f_{P_4}\leq 0.972$ | 78 | 64 | −155.41 | <800 Hz |
| $\lambda_2 \& \lambda_3$ | $f_{\lambda_2}\leq 0.007$ | $f_{\lambda_3}\leq 0.004$ | $f_{P_2}\leq 0.887$ | $f_{P_3}\leq 0.827$ | 78 | 62 | −154.08 | |
| $\lambda_2 \& \lambda_4$ | $f_{\lambda_2}\leq 0.011$ | $f_{\lambda_4}\leq 0.010$ | $f_{P_2}\leq 0.960$ | $f_{P_4}\leq 0.883$ | 78 | 62 | −154.94 | |
| $\lambda_3 \& \lambda_4$ | $f_{\lambda_3}\leq 0.002$ | $f_{\lambda_4}\leq 0.005$ | $f_{P_3}\leq 0.910$ | $f_{P_4}\leq 0.743$ | 78 | 64 | −155.07 | |

[a] Laser wavelengths in dual-wavelength operations same with them in single-wavelength operations as given in Fig. 5.

Note that, the switching among six dual-wavelength operations via adjusting the manual PCs was random, because the adjustment of two PCs was not quantitative. The only way to know which lasing state was achieved at a moment was to observe the output laser spectrum. However, by calibrating two PCs for every lasing mode beforehand, we also can switch one operating mode to another with certainty. Or we may use two programmable PCs (for instance, the voltage-controlled PCs, which can be adjusted by a digital or analog signal) instead of the manual PCs to achieve the switching with certainty in future.

**Table 2. Stability comparison of dual-wavelength lasing based on some typical techniques**

| Technique | Maximal Wavelength Fluctuation | Maximal power Fluctuation | Observation Time |
| --- | --- | --- | --- |
| CMFBG[a] + SA[b] [11] | N/A | 1.5 dB | 30 min |
| PM-FBG[c] + PM-CMFBG [22] | N/A | 1.5 dB | 30 min |
| HB-FBG + CRC [23] | 0.006 nm | 0.587 dB | 190 min |
| SI-HB-FBG + CRC [24] | 0.015 nm | 1.348 dB | 150 min |
| SI-FBGs + CRC [26] | 0.02 nm | 0.5 dB | 270 min |
| Two BPFs[d] + CRC [35] | 0.06 nm | 1.6 dB | 30 min |
| Structured PM-CFBG + SA [42] | 0.01 nm | 1 dB | N/A |
| Cascaded FBGs + CRC [43] | 0.01 nm | 2.54 dB | 2.5 h |
| Sagnac interferometer with FBG + SA [45] | 0.01 nm | 3 dB | 21 min |
| Sagnac interferometer with CFBG + SA [45] | Several pm | 1 dB | 20 min |
| **This work** | **0.014 nm** | **0.995** | **60 min** |

[a]CM-FBG: chirped Moiré fiber Bragg grating; [b]SA: saturable absorber; [c]PM-FBG: polarization maintaining FBG; [d]BPF: bandpass filter.

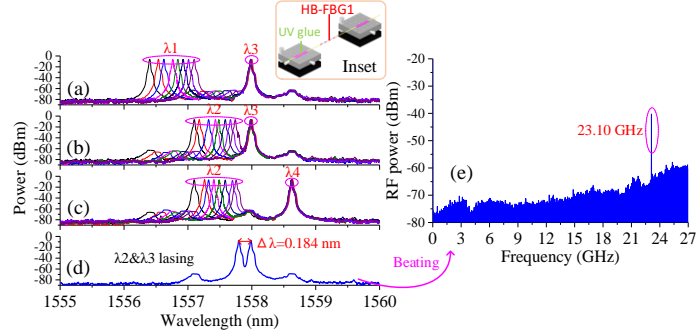

Fig. 13. Spectra of wavelength-spacing tuning for dual-wavelength operations lasing at (a) λ1&λ3, (b) λ2&λ3 and (c) λ2&λ4, respectively, through stretching HB-FBG1 using the setup shown in inset. (d) Spectrum of dual-wavelength lasing at λ2&λ3 with a spacing of 0.184 nm; (e) Photonic generation of microwave signal at 23.10 GHz through beating the dual-wavelength lasing output in (d), measured by ESA and high-speed PD.

The cascading design of HB-FBG1 and HB-FBG2 makes the tuning of wavelength spacing between the two lasers in a dual-wavelength operation possible, which may be used to fabricate a photonic generator of tunable microwave signals. The implementation method is to stretch the HB-FBG1 with two micro-displacement platforms as shown in the inset of Fig. 13. The fiber pigtails of the HB-FBG1 at both sides were affixed on the top-face of the platforms by the UV glue. In the operations lasing at λ1&λ3, λ1&λ4, λ2&λ3, and λ2&λ4, the wavelength spacing can be tuned when the HB-FBG1 is stretched by adjusting the micrometer of the platforms. Figures 13(a)–13(c) show the wavelength-spacing tuning for λ1&λ3, λ2&λ3, and λ2&λ4 lasing respectively. For demonstrating the photonic generation of microwave signal using the dual-wavelength lasing output, we could beat the two lasers in a high-speed PD. However, since the measureable frequency range of the ESA is only 26.5 GHz, the wavelength spacing of two lasers should be adjusted to less than 0.212 nm. Figure 13(d) shows the optical spectrum of the dual-wavelength lasing at λ2&λ3 with a spacing of 0.184 nm. By detecting the laser output with a 40 GHz PD to beat the two lasers, a microwave signal at 23.10 GHz was obtained as shown in Fig. 13(e), successfully demonstrating the feasibility and potential for photonic generation of tunable microwave signals using our 4WS-EDFL.

Based on the specially enhanced PHB effect formed in the EDF coiled in the TL-PC, mainly introduced by the PM-FCF, the switchable operations of power-equalized tri-wavelength and quad-wavelength lasing for our 4WS-EDFL were achieved through adjusting the PCs carefully. Figure 14 shows the typical spectra of four tri-wavelength operations respectively lasing at λ1&λ2&λ3, λ1&λ2&λ4, λ1&λ3&λ4 and λ2&λ3&λ4, and one four-wavelength operation. The OSNR is >79 dB for all operations, indicating the 4WS-EDFL was still with satisfying lasing performance in multi-wavelength operations. Similar with the demonstration in our previous work [24], the switching among 5 multi-wavelength operations was harder than that among 4 single- and 6 dual-wavelength operations for this 4WS-EDFL. However, the total switching difficulty was obviously lower than that in [24]. Since the stabilities in tri- and quad-wavelength operations are to be further improved to be comparable to those in single- and dual-wavelength operations, they will be characterized in detail once the improvement is completed.

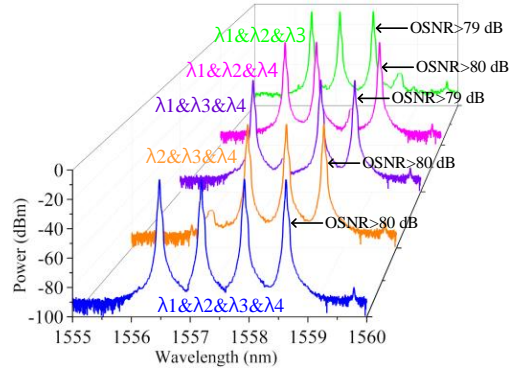

Fig. 14. Spectra of tri-wavelength operations lasing at λ1&λ2&λ3, λ1&λ2&λ4, λ1&λ3&λ4 and λ2&λ3&λ4 respectively and quad-wavelength operation lasing at λ1&λ2&λ3&λ4.

## 4. Conclusion

We have reported the first 4WS-EDFL enabled by a figure-8 compound-ring-cavity (F8-CRC) filter for SLM selection and a polarization-managed four-channel filter (PM-FCF) for defining lasing wavelengths. The F8-CRC is simply made of three couplers with extremely low cost and high filtering capability. The PM-FCF is structured by two HB-FBGs fusion-spliced together with a 45° polarization-axial-offset. We introduce a novel methodology utilizing the signal-flow graph combined with the Mason's rule to theoretically analyze a CRC filter in general, and apply it to obtain the important design parameters for the F8-CRC used in the proposed 4WS-EDFL. This method can be a common approach for designing, fabricating and characterizing CRC filters aiming for applying in SLM fiber lasers. The measured transmission spectrum of the fabricated F8-CRC is highly consistent with the simulated result. By integrating the F8-CRC and the PM-FCF in a fiber laser theoretical model, we validate that the SLM lasing can be achieved in each channel of the PM-FCF. Due to the use of the PM-FCF, the polarization ellipses' major axes of any two lasing wavelengths are 45° or an integer multiple of 45° with respect to each other. The PHB effect in the EDF coiled in the three-loop PC is significantly enhanced, compared to that in our previous work [24]. By adjusting the two PCs, the switching among 15 lasing states was experimentally validated, including 4 single-, 6 dual-, 4 tri- and 1 quad-wavelength operations. In single-wavelength operations, the 4 lasing outputs are all in stable SLM oscillation, typically with a linewidth of <600 Hz, a RIN of ≤−154.58 dB/Hz@≥3 MHz and an output power fluctuation of ≤±3.45%. In dual-wavelength operations, all 6 output states have very similar performances, also with the performance parameters close to those of single-wavelength operations. In addition, the switching among tri- and quad-wavelength operations is much easier than that in [24], which is validated during the experiments. We also demonstrate the wavelength spacing adjustment of the dual-wavelength operations, which can be used for photonic generation of tunable microwave signals, and obtain a 23.10 GHz signal successfully by beating the dual-wavelength lasing with a spacing of 0.184 nm. We believe that the outstanding performances of our 4WS-EDFL can be further improved if special temperature compensation and vibration isolation are employed in future for practical applications.

**Funding.** National Natural Science Foundation of China (61975049, 61775128, 61827818, 12004092); Hebei Provincial Natural Science Foundation for Outstanding Young Scholars (F2020201001).

**Disclosures.** The authors declare no conflicts of interest.

**Data availability.** Data underlying the results presented in this paper are not publicly available at this time but may be obtained from the authors upon reasonable request.